\documentstyle[preprint,aps]{revtex}

\begin{document}
\draft
\title{A one-dimensional lattice model for a 
quantum mechanical free particle}
\author{A. C. de la Torre }
\address{Departamento de F\'{\i}sica,
 Universidad Nacional de Mar del Plata,\\
 Funes 3350, 7600 Mar del Plata, Argentina.\\
dltorre@mdp.edu.ar}
\date{\today}
\maketitle
\begin{abstract}
Two types of particles, $A$ and $B$ with their corresponding antiparticles,
 are defined in a one dimensional cyclic lattice with an odd number of sites.
 In each step of time evolution, 
each particle acts as  a source for the polarization field of the other 
type of particle with  nonlocal action but with an effect decreasing with the 
distance: 
$A \rightarrow \cdots \bar{B} B \bar{B}  B \bar{B} \cdots$ ;
$B \rightarrow \cdots  A  \bar{A}  A \bar{A}  A   \cdots$. 
It is shown that the combined distribution of these particles 
obeys the time evolution of a free particle as given by quantum mechanics. 
\end{abstract}
\narrowtext
The modeling of physical reality by means of fictitious particles 
that move and react in a substrate of different geometrical structures 
has been a fruitful strategy that has extended our analysis capabilities 
beyond the domain associated with differential equations\cite{model}. 
The particles
involved in these models are classical in the sense that they are given 
precise location and velocity. This is clearly inadequate for the 
modeling of quantum systems that require, not only the indeterminacies 
imposed by  Heisenberg's principle, but also nonlocal correlations 
between {\em commuting} observables,  suggested by the Einstein 
Podosky Rosen argument\cite{epr} and  empirically 
established in the violation of Bell inequalities\cite{bell,exp}. 
However this does not forbids the modeling of quantum systems if we 
{\em do not} identify the particles of the model with the quantum particles. 
It is possible, as will be seen in this work, to associate the real 
quantum particle to a combined distribution of two types of fictitious 
particles with nonlocal interaction. This simple example model can be 
trivially extended to higher dimensions of space and to a higher number 
of noninteracting quantum particles and it
 provides a new point of view to study the peculiarities of 
quantum mechanics.
\par
Let us assume a one dimensional lattice with $N$ sites on a circle and 
lattice constant $a$. We assume $N$ to be an {\em odd} integer. The 
reason for this restriction will become clear later. The inclusion of 
{\em even} values for $N$ would introduce unwanted complications in the model.
Each site can be occupied by any number of particles of type $A$, $B$ 
or by their corresponding antiparticles $\bar{A}$, $\bar{B}$. Particles and 
antiparticles of the same type annihilate in each site of the lattice 
leaving only the remaining excess of particles or antiparticles of both 
types $A$ and $B$. At each time step, $t\rightarrow t+1$, corresponding 
to a time evolution by a small amount $\tau$, the particles of type $A$ 
create  antiparticles $\bar{B}$ in the same site, particles $B$ in the 
first neighboring sites, $\bar{B}$ in the second neighboring sites and 
so on. In a similar way, particles $B$ create particles $A$ and $\bar{A}$.
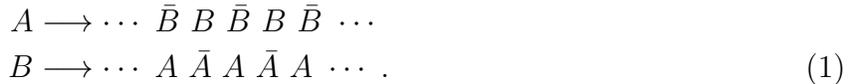
\begin{eqnarray}
A &\longrightarrow& \cdots\  \bar{B}\  B\  \bar{B}\  B\  \bar{B}\  \cdots
\nonumber \\
B &\longrightarrow& \cdots\  A\  \bar{A}\  A\  \bar{A}\  A\   \cdots
\ .
\end{eqnarray}
The same reactions occur exchanging particles and antiparticles. 
This creation process extends to the right and left of each site  
up to the two opposing sites in the circle. Since $N$ is odd, in these two 
sites particles of the same sign (either particles 
or antiparticles)  are created.
The {\em number} of particles or antiparticles created decreases 
with the distance $d$ roughly like $1/d^{2}$ for a distribution of 
particles confined in a small region within a large lattice as will 
be precisely 
stated later. Before we write the master equation for the time evolution,
we can notice some qualitative features of the process. It is easy to see 
that the process has diffusion. If we start, for instance, with 
some number of $A$ particles in one site, after {\em two} time steps, 
some $\bar{A}$ antiparticles have been created at the same site reducing 
the number of $A$ particles, but also 
some $A$ particles appear in the first neighboring sites. The net effect is 
diffusion. It is less obvious that, even though the process has left-right 
symmetry, we may also have drift to the right or to the left.
In order to see how this is possible we notice that  $A$ particles
{\em reject} $B$ particles from the site because $\bar{B}$ are created there, 
whereas $B$ particles {\em attract} neighboring $A$ particles to his site. 
Therefore if we have an asymmetric configuration like $AB$, the center of 
the combined distribution will move towards $B$. The drift direction and 
velocity is then {\em encoded} in the shape and relative distribution of 
both types of particles. We will see  that, although the distribution 
of particles are widely distorted after few time steps, the drift 
direction and velocity remain invariant. 
\par
A convenient way to label the sites of the circular lattice is 
by an index
$s$ running from $-L$ to $L$. Since  the number of sites $N=2L+1$ is odd,
the index $s$ will be integer. It is of course irrelevant which site 
has the label $s=0$.  Let $a_{s}(t)$ and $b_{s}(t)$ be the number of 
particles of type $A$ and $B$ 
respectively at the site $s$ at time $t$, normalized in a way that will 
be specified later (anyway the master equation is independent of the 
normalization). When $a_{s}(t)$ or $b_{s}(t)$ take negative values they 
denote the number of {\em antiparticles}. At a particular site of the 
lattice, the number of particles change as particles or antiparticles 
are created in it by the particles in other sites. The time evolution 
of the process is then defined by the equations
\begin{eqnarray}
a_{s}(t+1) &=& a_{s}(t)+\tau g^{2} \sum_{d=-L}^{L} b_{[s+d]}(t) F(d)
\nonumber \\
b_{s}(t+1) &=& b_{s}(t)-\tau g^{2} \sum_{d=-L}^{L} a_{[s+d]}(t) F(d)
\ ,
\end{eqnarray}
where the square brackets in the index, $[s+d]$, denotes ``modulo $N$'', 
that is, with a value in the closed interval $[-L,L]$;  $g$ is related 
to the lattice constant $a$ by  $g=(2\pi)/(Na)$ (it 
corresponds to the reciprocal lattice constant); $\tau$ is a time scale
 small enough to make $\tau g^{2} N^{2}\ll 1$, and the function 
of the distance $F(d)$ is defined as
\begin{equation}
F(d) = \frac{1}{N} \sum_{k=-L}^{L}k^{2} e^{i\frac{2\pi}{N} kd}=
\left\{ \begin{array}{ll}
          (-1)^{d}\frac{\cos(\pi d/N)}{2\sin^{2}(\pi d/N)} 
           & \mbox{ if $d=\pm 1, \pm 2,..., \pm L$} \\
          \frac{1}{12} (N^{2}-1)
        & \mbox{ if $d=0$}  
         \end{array} \right. \ .
\end{equation}
For later use we define a similar function $G(d)$ as:
\begin{equation}
G(d) = \frac{i}{N} \sum_{k=-L}^{L}k e^{i\frac{2\pi}{N} kd}=
\left\{ \begin{array}{ll}
          \frac{(-1)^{d}}{2 \sin(\pi d/N)} 
           & \mbox{ if $d=\pm 1, \pm 2,..., \pm 2L$} \\
          0
        & \mbox{ if $d=0$}  
         \end{array} \right. \ .
\end{equation}
The alternating sign in the definition of $F(d)$ corresponds to the fact 
that particles and antiparticles are created at alternating sites, and 
the different sign in Eq.2 is due to the difference in the r\^ole of 
particle and antiparticle in Relation 1.  If the particles are confined 
in a 
small region within a large lattice, the main contribution in the sums 
of Eq.2 comes from terms with distance $|d| \ll N$. In this limit we 
have $|F(d)|\approx 1/d^{2}$ as mentioned before. 
\par
The number of $A$ or $B$ particles 
are not conserved in the time evolution. Neither is  the 
sum of particles conserved. A quantity that is conserved in the time 
evolution of the process is the sum of 
the square of the number of $A$ particles (or antiparticles) plus the sum of 
the square of the number of   $B$ particles (or antiparticles). This 
conserved quantity can be used for normalization and can be given a 
physical meaning like energy density or probability density. It is 
therefore relevant to define a {\em combined distribution} 
$a_{s}^{2}(t)+b_{s}^{2}(t)$, associated to this 
conserved quantity. We will see that the drift velocity of 
this combined distribution, given by
\begin{equation}
\langle V \rangle=4 g \sum_{s , r}  a_{s}(t) b_{r}(t)
          G(s-r) 
          \ ,
\end{equation}
is also conserved in the time evolution. Given a distribution
$\{a_{s}(t),b_{s}(t)\}$, we can change the drift velocity by an amount
$v$, {\em without} changing the shape of the combined distribution by means
of the local transformation
\begin{eqnarray}
a'_{s}(t) &=& a_{s}(t) \cos(vas/2) - b_{s}(t) \sin(vas/2)
\nonumber \\
b'_{s}(t) &=& a_{s}(t) \sin(vas/2) + b_{s}(t)\cos(vas/2)
\ .
\end{eqnarray}
\par
These features have been checked in a computer simulation of the process.
A circular lattice with $N=801$ sites ($L=400$)  and with lattice constant 
$a=1$ was 
chosen. Several shapes of initial distributions were tried: gaussian, 
uniform and random, with several widths and drift velocities.  The time
dependence of $M(t)=\sum_{s} (a_{s}^{2}(t)+b_{s}^{2}(t) )$ and of the 
drift velocity given in Eq.5 was studied. Taking a time step $\tau=10^{-3}$,
we found that these quantities remain constant after $t=1000$ time steps, 
with a relative variation less than $10^{-5}$ for the gaussian case, 
$4\times 10^{-4}$ for the uniform distribution, and $0.04$ for the random 
distribution. For larger time steps, $\tau=0.005$, these quantities 
remain constant (less than $1\%$ relative variation) for the gaussian 
and uniform case at $t=1000$ but the random case begins to show 
significant departure from constancy. At $\tau=0.010$ only in the 
gaussian case these quantities remain constant (less than $0.1\%$).
The time evolution of the shape of the combined distribution strongly 
reminds the time evolution of quantum mechanical wave packets. For 
instance, a gaussian distribution for $A$ and $B$ particles, modified 
by Eq.6 in order to have drift, will evolve increasing the width and 
drifting but maintaining the gaussian shape. A uniform distribution 
will develop  side lobes in the evolution. A  remarkable feature
is that the process smoothens out the random fluctuations
of  an initial distribution. 
\par
The resemblance of the process with quantum mechanics is striking.
 We will indeed show that the process here defined 
corresponds to a quantum mechanical free particle in a lattice. 
Let us define then an $N$ dimensional Hilbert space spanned by a 
basis $\{\varphi_{s}\}$ $s=-L, -L+1, \cdots,L$  corresponding to the 
eigenvectors  of the position operator $X$. Then, 
$X \varphi_{s}=as\varphi_{s}$. In this finite dimensional Hilbert space, 
we can not define the momentum operator $P$ by means of the usual 
commutation relation. The alternative way to define $P$ is to choose 
first an {\em unbiased basis}\cite{woo,per,dlt}  $\{\phi _{k}\}$, 
\begin{equation}
\phi _{k} = \frac{1}{\sqrt{N}} \sum_{s=-L}^{L} e^{i\frac{2\pi}{N} k s} 
 \varphi_{s}\ ,
\end{equation}
and with it, we define the momentum by the spectral decomposition
\begin{eqnarray}
P &=& 
\sum_{k=-L}^{L} g k \phi_{k}\langle\phi_{k},\cdot\rangle
\nonumber\\
 &=& \frac{1}{N} \sum_{k,s,r} g k 
e^{i\frac{2\pi}{N} k (s-r)} 
 \varphi_{s}\langle\varphi_{r},\cdot\rangle
\ . 
\end{eqnarray}
The momentum eigenvalues and the relative phases to build the basis 
$\{\phi_{k}\}$ have been chosen such that $P$ is the generator 
of translations. That is, with this choice, the operator 
$U_{a}= \exp(-iaP)$ is such that $U_{a}\varphi_{s}=\varphi_{s+1}$.
The translation is cyclic at the border, 
$U_{a}\varphi_{L}=\varphi_{-L}$. 
If we had taken $N$ even, the right hand side of this equation should 
have a minus sign. This would complicate the model of Relation 1 introducing 
a change of sign at some appropriated places. In order to have a simple 
lattice model for the quantum free particle we prefer to restrict 
ourselves to odd values of $N$. 
\par
The state of a free quantum particle, given by
\begin{equation}
\Psi(t) = \sum_{s=-L}^{L} c_{s}(t) \varphi_{s}\ ,
\end{equation}
will change  according to the  time
evolution operator  (we set $\hbar=2m=1$)
\begin{equation}
U_{t} = \exp(-iP^{2}t)\ .
\end{equation}
Let us consider the evolution of the coefficients of the 
 expansion given in Eq.9, in one step of discretized time: $t_{0}=\tau t$
and $t_{1}=\tau (t+1)$ with a small time scale 
$\tau$ and $t$ positive integer.  We have
\begin{equation}
c_{s}(t+1) = \sum_{r=-L}^{L} c_{r}(t) 
\langle\varphi_{s},U_{\tau}\varphi_{r}\rangle\ .
\end{equation}
For $\tau$ small enough such that $\tau \|P^{2}\| \ll 1$, that 
is $\tau \ll (a/\pi)^{2}$, the time evolution operator can be linearized
and we obtain
\begin{equation}
c_{s}(t+1) =c_{s}(t)-i\tau\sum_{r=-L}^{L} c_{r}(t) 
\langle\varphi_{s},P^{2}\varphi_{r}\rangle\ .
\end{equation}
Using Eq.8 we calculate the matrix element
\begin{equation}
\langle\varphi_{s},P^{2}\varphi_{r}\rangle=g^{2}
 \frac{1}{N} \sum_{k=-L}^{L}k^{2} e^{i\frac{2\pi}{N} k (s-r)}
\ .
\end{equation}
 We have then
\begin{equation}
c_{s}(t+1) = c_{s}(t)-i\tau g^{2} \sum_{r=-L}^{L} c_{r}(t) F(s-r)
\ .
\end{equation}
Reordering the terms in the sum and using the ``modulo $N$'' notation, 
we get
\begin{equation}
c_{s}(t+1) = c_{s}(t)-i\tau g^{2} \sum_{d=-L}^{L} c_{[s+d]}(t) F(d)
\ .
\end{equation}
Finally if we explicitly write the coefficients with real and 
imaginary part, $c_{s}(t)=a_{s}(t)+ib_{s}(t)$, we get Eq.2 above. 
\par
We can here check that $M(t)=\sum_{s} |c_{s}(t)|^{2}$ is conserved. 
\begin{equation}
M(t+1) = M(t)+\tau^{2} g^{4} \sum_{r,u} c_{r}(t)c_{u}^{*}(t) 
\sum_{s}F(r-s)F(s-u)
\ .
\end{equation}
The term linear in $\tau$ vanishes because the symmetric function 
$F$ appears multiplied by an anti-symmetric factor. We see here that 
the ``derivative'' $(M(t+1) - M(t))/\tau$ vanishes
like $\tau$ in agreement with the numerical simulation of the process.
We can write the functions $F$ in their summation representations 
 and, performing the sum over $s$,  we get
\begin{equation}
\sum_{s}F(r-s)F(s-u)= \frac{1}{N} \sum_{k=-L}^{L}k^{4} 
e^{i\frac{2\pi}{N} k (r-u)}
\ .
\end{equation}
This sum can be evaluated as was done in Eq.3 and 4 but we don't need it.
 In the limit $N\gg d=r-u$, that is, when the 
particles are confined in a small region of a large lattice, we get
\begin{equation}
M(t+1) = M(t)+\tau^{2} \frac{\pi^{4}}{5}\left( 
2 \sum_{r\neq u} c_{r}(t)c_{u}^{*}(t)\frac{(-1)^{(r-u)}}{(r-u)^{2}}
+ M(t) \right)
\ .
\end{equation}
 A similar result is obtained for the drift velocity, 
proportional to the expectation value of $P$.
\begin{equation}
\langle P \rangle_{t} = -i g \sum_{s , r} c_{s}^{*}(t)c_{r}(t)
G(s-r)
\ .
\end{equation}
In terms of $a_{s}(t)$ and $b_{s}(t)$  
this equation becomes the Eq.5 above. 
Here again, considering the time evolution $\langle P \rangle_{t+1}$, 
the term linear in $\tau$ vanishes because it contains
$\sum_{s}[F(u-s)G(s-r)-G(u-s)F(s-r)]$ which is zero as can be calculated 
 with the summation representation of the functions $F$ and $G$.
We obtain then
\begin{equation}
\langle P \rangle_{t+1} = \langle P \rangle_{t} -i \tau^{2} g^{5}
 \sum_{u,v}c_{u}^{*}(t) c_{v}(t)
\sum_{s,r}F(u-s)G(s-r)F(r-v) 
\ .
\end{equation}
showing that the drift velocity is constant to order $\tau$, that is, 
the ``derivative'' vanishes with $\tau$ in agreement with the 
numerical simulation of the process. 
Finally, applying a boost transformation $\exp( iXv/2)$ to the state
of Eq.9, we prove equation (6).
\par
The one dimensional lattice model here presented provides a simple 
representation  for the position and momentum  of a free quantum 
mechanical particle. 
In this model we require that $N$ should be odd. Let us see what happens 
in the 
case where $N$ is even. In this case, the model evolves according to 
Eq.2 with the summations running from $-N/2$ to $N/2$ and with the 
same function $F(d)$ defined in Eq.3. Notice that this function vanishes 
at the extreme values of $d$, that is $F(\pm N/2)=0$. This model can be 
interesting in itself but it is no longer equivalent to the quantum 
mechanical system. The connection is lost in the step from Eq.14 to Eq.15. 
For the cases when the argument $s-r$ of the function $F$ in Eq.14 take 
values exceeding $N/2$, we should introduce a minus sign if we want to 
change the argument to $d$ as in Eq.15 (in the case $N$ odd, no sign 
change is needed).  The reason for this change can be 
traced to the change in sign produced by the translation operator 
when the site labeled by $\pm L$ is crossed as mentioned after Eq.8.
It would be possible to include even values for $N$ but at the cost of 
complicating the model. For this we would have to change the rules 
of Relation 1 exchanging particles and antiparticles when we cross 
the site with label $\pm L$. These complications are 
unwanted and we prefer to accept the fact that  position and 
momentum of a quantum mechanical particle can be easily modeled only with 
a cyclic lattice with an odd number of sites. 
In the case of a quantum 
particle confined in a very small region (say, 10 sites) of a very 
large lattice (say, close to one million sites) it doesn't matter
whether $N$ is even or odd for all times until, due to drift or diffusion, 
the distribution reaches the sites with label close to $\pm N/2$. However 
for small lattices and for extended distributions it does matter, 
and only in the odd $N$ case the model of Relation 1 describes a 
quantum mechanical particle. This is a further indication 
of the essential nonlocal character of quantum mechanics. There is another 
case in quantum mechanics where an even or odd number of states has 
important cualitative 
consequences. This is in finite dimensional realizations of 
angular momentum. Whereas {\em intrinsic} angular momentum, the spin, of a 
particle can have an even or odd number of states, the {\em orbital} 
angular momentum, arising from position and momentum, can only have 
a realization with an odd number of states.  
\par
The model presented can be extended from the free particle to the case 
of a position dependent potential. The general structure of the process 
shown in Relation 1 remains unchanged but the function $F$ of Eq.2 will 
not be 
given by Eq.3 but will have to be calculated from an appropriate 
time evolution operator. The process can also be extended to two or three
space dimensions but with larger computer requirements for the numerical 
simulations.
\par
Since the advent of quantum mechanics, there have been numerous attempts 
to develop a classical image for quantum behavior. For the reasons already
mentioned at the beginning, the attempts in terms of 
{\em particles} are doomed. 
The model here presented, suggests the possibility of a classical image 
in terms of two {\em fields}  $A$ and $B$ where each field acts as a 
source for the polarization of the other. As happens with the 
electromagnetic fields, the energy, or whatever conserved quantity, is 
given by the sum of the square of both fields. The consideration of these
field may provide a new point of view for studying the peculiarities of 
quantum mechanics.
\par
I would like to thank H. M\'artin and A. Daleo  for discussions and comments.
This work has been done with partial support from  ``Consejo 
Nacional de Investigaciones Cient\'{\i}ficas y T\'ecnicas'' 
(CONICET), Argentina.

\end{document}